\documentclass[letter]{aa}  

\usepackage{graphicx}
\usepackage{natbib}
\usepackage{txfonts}
\usepackage{hyperref}

\usepackage{orcidlink}
\usepackage{xcolor}

\newcommand{\bluebf}[1]{\textcolor{blue}{\textbf{#1}}}
\renewcommand{\bluebf}[1]{#1}

\begin{document}

   \title{EK Cha b: multi-tracer detection of a protoplanet candidate in a gas-rich environment.}
   \titlerunning{EK Cha b: The protoplanet candidate}

\author{
  Felipe Alarcón\inst{1}
  \orcidlink{0000-0002-2692-7862}
  \and C. Swastik\inst{1}
  \orcidlink{0000-0003-1371-8890}
  \and Stefano Facchini\inst{1}
  \orcidlink{0000-0003-4689-2684}
  \and Edwin A. Bergin\inst{2}
  \orcidlink{0000-0003-4179-6394}
  \and Gabriele Cugno\inst{3}
  \orcidlink{0000-0001-7255-3251}
}

\institute{
  Dipartimento di Fisica, Università degli Studi di Milano,
  Via Celoria 16, 20133 Milano, Italy
  \and
  Department of Astronomy, University of Michigan,
  1085 South University Avenue, Ann Arbor, MI 48109, USA
  \and
  Department of Astrophysics, University of Z\"urich,
  Winterthurerstrasse 190, 8057 Z\"urich, Switzerland
}

   \date{}

 
  \abstract
   {Despite sustained community efforts and advances in instrumentation and high-contrast imaging, the search for protoplanets remains difficult, with only two widely accepted systems to date in Class II systems (PDS 70 and WISPIT 2). Here we present the detection of EK Cha b, a new protoplanet candidate, identified through CO emission, millimeter continuum and 0.82 $\mu$m broadband photometry, linked to EK Cha, a young late-type M star.}
   {The detection of a new protoplanet candidate offers an additional opportunity to advance our understanding of planet formation. By exploiting the wealth of archival data available for nearby star-forming regions, we aim to place EK Cha b in context, characterize its properties, and motivate future campaigns targeting similar systems.}
   {We analyzed high-spatial-resolution archival observations of EK Cha from HST, VLT/NACO, VLT/MUSE, and ALMA. We searched for EK Cha b in broadband continuum and gas tracers, reporting the measured flux with uncertainties, and upper limits for non-detections.}
   {EK Cha b shows a spectrally-resolved footprint that matches the CO J=2--1 emission, together with 0.82 $\mu$m and millimeter continuum emission. EK Cha b is spectrally resolved in CO J=2--1 emission, kinematically consistent with a circumplanetary disk. Best fits from evolutionary models place EK Cha b as a planetary mass companion, although deeper observations are required to fully confirm its nature.
   }
   {Although giant planets are uncommon around M dwarfs, these stars host the highest planet occurrence rate while remaining severely underrepresented in protoplanet searches. EK Cha b demonstrates that such systems are now within reach, and the advent of extremely large telescopes and improved high-contrast imaging — where the lower host luminosity relaxes the required contrast — will open this overlooked regime to systematic exploration.}


   \maketitle
%
\section{Introduction}\label{Sec: Intro}

Advances in High-Contrast Imaging (HCI) have enabled the first direct detections of forming planets. However, only two Class II systems so far show embedded, imaged protoplanets with general consensus: PDS 70 \citep{Keppler..2018, Haffert..2019}, and WISPIT 2 \citep{WISPIT2, Close..Wispit, Lawlor}. To date, only PDS 70 c \citep{Benisty..21} has a millimeter detection of a circumplanetary disk (CPD). Planetary-mass companions (PMCs) at wide orbits have shown infrared CPD emission \bluebf{(e.g., Delorme 1 AB b \cite{Malin..2025}; GQ Lup b \cite{Cugno..2024..GQLup}, among others)}, although millimeter emission through deep ALMA observations has only been detected in SR 12 c \citep{Wu..2022}.


HCI campaigns of protoplanetary disks have mostly focused on disks with bright millimeter emission, prioritized by early ALMA programs, and disks around luminous host stars with  good adaptive optics (AO) performance. Both choices introduce biases. Dust-rich disks can produce strong extinction in embedded planets \citep{Alarcon..ext, Cugno..2025}, while the higher luminosity of the hosts implies larger planet-to-star contrasts. As a result, faint and compact late-type stars are systematically overlooked. Although giant planets are less common on wide orbits around low-mass stars \citep{Vigan..2021}, M stars still present the highest overall planet occurrence rate \citep{Yiang..2020, Hsu..2020}. Given the advances in HCI and AO, the advent of the ELT makes the inner regions of late-type stars an increasingly accessible — and underexplored — frontier for protoplanet detection.

The $\eta$ Cha association provides a nearby laboratory for such searches. At an age of $t\sim8\pm1$ Myr and with $23$ predominantly low-mass members \citep{Roccatagliata..2024}, and it has been observed extensively across the electromagnetic spectrum \citep[e.g. ][]{Riviere..2015, Megeathh..2005, Sicilia-Aguilar..2009, Rugel..2018, Mamajek..1999}. Nevertheless, the faintness and compactness of its members have left them comparatively unexplored in contemporary high angular resolution observations.

Part of the $\eta$ Cha association is the accreting M dwarf EK Cha (RECX 5) \bluebf{with a singular spectral energy distribution (SED) \citep{Bouwman..2006}.} Here, we report a protoplanet candidate, EK Cha b, at a projected separation of 46 au. EK Cha b is detected in kinematically resolved CO emission, millimeter dust continuum, and broadband far-red/near-infrared imaging. EK Cha b may be the first protoplanet candidate with a kinematically resolved CPD detected in millimeter emission. At 98.2$^{+0.15}_{-0.25}$ pc \citep{Gaia..DR3}, with negligible foreground extinction \citep{Lyo..06}, EK Cha offers a uniquely favorable, underexplored window onto giant-planet formation.

This letter is organized as follows. In Section~\ref{Sec: Data} we present the datasets and the methodology. In Section~\ref{Sec: Results} we report the detection of EK Cha b, the protoplanet candidate through three independent tracers. Finally, in Section~\ref{Sec: Discussion} we conclude and discuss possible avenues for protoplanet discovery.

\section{Methods and Data}\label{Sec: Data}

\subsection{Multi-wavelength EK Cha archival observations}

We performed an archival search for data on EK Cha spanning multiple regions of the electromagnetic spectrum. We selected those datasets providing sufficient spatial resolution to disentangle the candidate companion that was first observed in the ALMA 1.3 mm data. We thus analyze the following datasets: (1) ALMA Band 6 (1.3 mm) and Band 7 (0.8~mm) observations, including the millimeter dust continuum and the CO J=2--1 and CO J=3--2 lines. (2) optical photometry from HST/ACS in the F606W (0.606 $\mu$m) and F814W (0.82 $\mu$m) filters; (3) ground-based near-infrared observations from VLT/NACO in the \textit{H} (1.6 $\mu$m) and \textit{K$_s$} (2.2 $\mu$m) bands; and (4) MUSE-NFM integral-field-unit (IFU) observations, including the H$\alpha$ and H$\beta$ accretion lines. Details of the sensitivity, data reduction, and on-source times for each observation are provided in Appendix~\ref{App: Data}.   

\begin{figure}[h]
\includegraphics[width=0.90\linewidth]{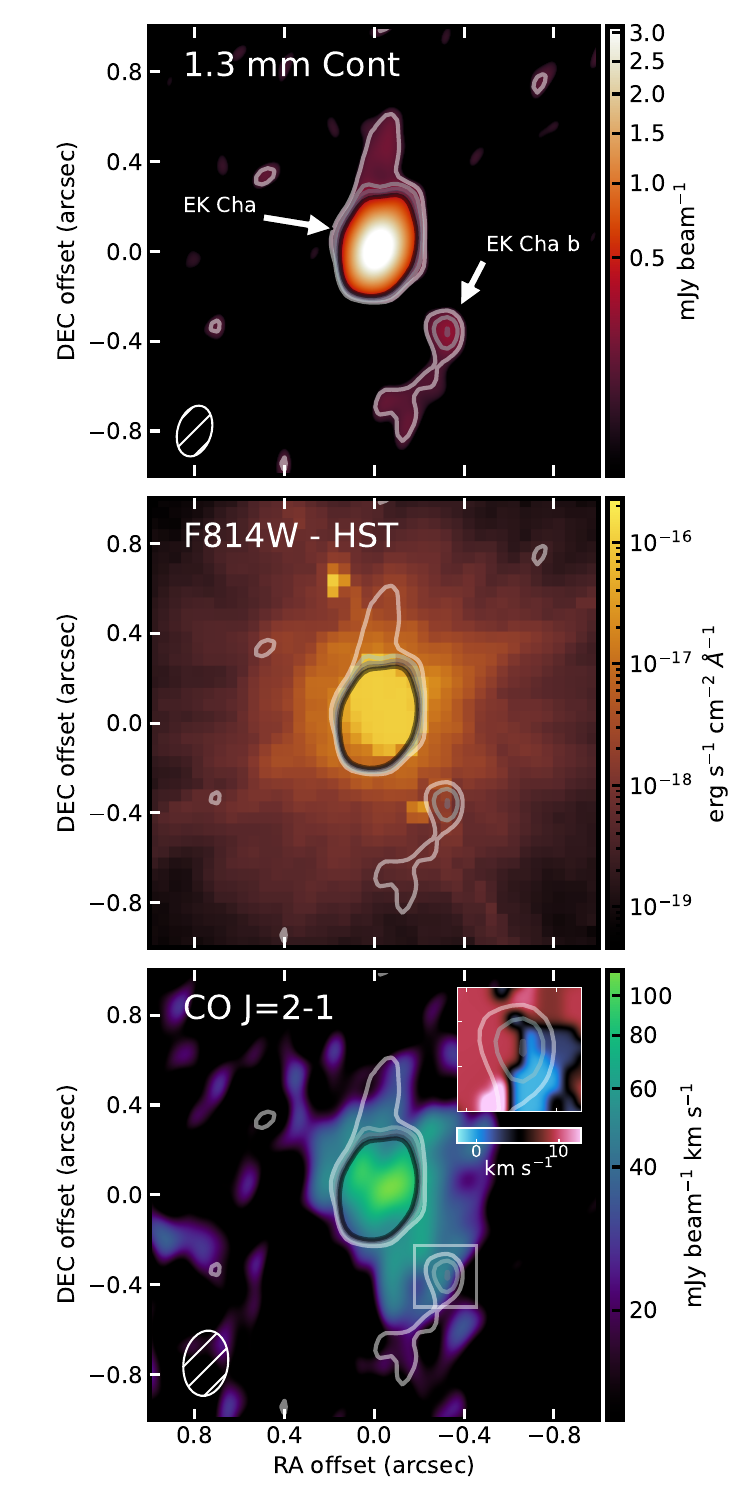}
\caption{\textbf{Top}: ALMA dust continuum 1.3 mm emission of EK Cha, showing a detected unresolved source to the SW (5.1$\sigma$). \textbf{Middle}: HST/ACS F814W photometric image (0.82 $\mu$m) with \bluebf{a synthetic PSF's FWHM of $\approx$84 mas} overlaid with 1.3 mm emission contours showing the relative location of EK Cha b. \textbf{Bottom}: Velocity-integrated flux density map of the CO J=2--1 line overlaid with ALMA 1.3 mm contours. The CO emission at EK Cha b reaches a  5.3$\sigma$ significance. The inset shows the CO J=2--1  velocity centroid in EK Cha b, with a gradient that is consistent with a keplerian disk. The beams of the ALMA data are shown on the bottom-left of the respective panels. 
}
\label{Fig:Gallery}
\end{figure} 

The imaging observations, regardless of whether they yield a detection, allow us to place stringent constraints on the EK Cha b's emission. We use these constraints to characterize EK Cha b through an SED fitting with atmospheric evolution models, which is further detailed in Appendix \ref{App: SED}.

\subsection{Fitting the ALMA visibilities with \texttt{galario} and \texttt{frank}}

Even though EK Cha's properties, which we list in Table \ref{Tab: EKCHa_pars},  have been characterized  through multiple observations, resolved observations have not been reported. Thus, we used the ALMA Band 6 observations to measure the disk's phase-center offset, the disk's inclination, and  the position angle  by fitting a 2D Gaussian to the visibilities using the \texttt{galario} package \citep{galario}. Then, we performed a non-parametric modeling of EK Cha's millimeter continuum visibilities with the \texttt{frank} package \citep{Jennings} to search for radial substructures. \texttt{frank} increases the effective spatial resolution by collapsing one of the spatial dimensions by assuming azimuthal symmetry. We detail the specific details of the visibility fitting in Appendix \ref{App: Frank}.

\section{Results}\label{Sec: Results}

\subsection{The detection of EK Cha b with HST and ALMA}

  In Fig. \ref{Fig:Gallery} we show the detection of a protoplanet candidate in EK Cha, EK Cha b, through three independent tracers: (1) In broad, spectrally resolved, CO J=2--1 line emission, hinting at the gas presence around a localized compact strong gravitational potential, with an integrated emission detected at 5.3$\sigma$. (2) In ALMA 1.3 mm dust continuum, detached from the primary disk \bluebf{(5.1$\sigma$)}, tracing EK Cha b's CPD. (3) A photometric detection ($\sim$ 300$\sigma$) in HST observations with the F814W filter (0.82 $\mu$m).  We also detect the emission of CO J=3--2; however, the low spatial resolution of that dataset prevents us from claiming a clear resolved detection.

\begin{figure}
    \includegraphics[width=0.5\textwidth]{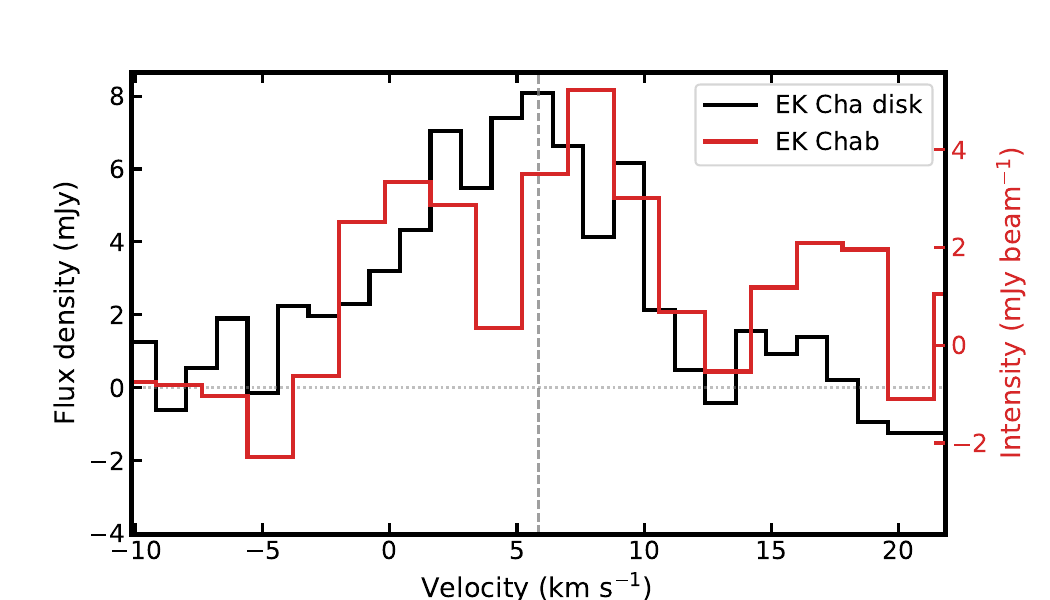} 
\caption{CO J=2--1 spectrum at both the primary disk and at the location of EK Cha b as detected in the continuum. The comoving emission is key to link the emission to the same system and to not associate it with a background object. The double-peaked velocity-broadened emission of EK Cha b suggests a collapsing structure, or one rotating on its own instead of a clump orbiting EK Cha.}
\label{Fig 2: CO}
\end{figure}

We show EK Cha b's spectrally resolved CO emission in Fig. \ref{Fig 2: CO}. The spectrum extracted over one beam centered on the CPD (red) shows broad CO emission centered at $4.3\pm 1.8$ km s$^{-1}$, around EK Cha's systemic velocity (5.83 km s$^{-1}$), consistent with co-motion from the spatially integrated EK Cha disk emission in black. It is serendipitously located close to the disk minor axis ($\Delta$PA$\sim$20\degr), so a small systemic velocity offset of EK Cha b's systemic velocity is within the expected values for Keplerian rotation around EK Cha. Furthermore, the broad emission discards an isolated clump rotating around the primary, but rather a strong localized gravitational potential deep enough to be consistent with gas collapsing, or with a Keplerian rotation of its own, such as the one expected in a CPD.
  
We measure EK Cha b’s separation from the offset between its ALMA 1.3 mm emission peak and the disk center inferred through \texttt{galario} visibility modeling. Using the astrometric uncertainty based on the ALMA Technical Handbook \citep[Sec. 10.5.2]{ALMA_techbook}, we obtain 461 $\pm$ 42 mas (46 $\pm$ 4 au). EK Cha b lies beyond the millimeter continuum disk (r$\leq$8) au but remains embedded in the gaseous disk (see Fig \ref{Fig:Gallery}). Archival HST images detect it in the F814W filter but not in the visible F606W filter, shown in Appendix \ref{App: Data}.

The NACO non-detections place EK Cha b below the brown-dwarf mass regime (Appendix \ref{App: VLT}). Combining infrared and optical observations, the SED can be explained by a PMC with an effective temperature of 2116$^{+118}_{-187}$ K, a radius of 1.7$^{+0.4}_{-0.4}$ $R_{\rm Jup}$, a mass of 8.50$^{+6.46}_{-3.92}$  $M_{\rm Jup}$ and an extinction of $A_V=5.0^{+1.0}_{-0.9}$ mag (see Fig. \ref{Fig SED}). The atmospheric fitting placing EK Cha below the stellar regime is mostly constrained by the F814W detection and photometric upper limits, despite the CO  spectroscopic detection and millimeter continuum emission. Thus, more photometric points and at deeper sensitivity are required for a more constraining characterization of EK Cha b and its potential CPD. 

\begin{table}
\caption{Stellar and disk parameters of EK Cha (RECX 5).}
\label{Tab: EKCHa_pars}
\centering
\setlength{\tabcolsep}{3pt}
\begin{tabular}{@{}ll|ll@{}}
\hline\hline
Parameter & Value & Parameter & Value \\
\hline
Spectral type & M5$^{(1)}$
  & Age & $8\pm1$ Myr$^{(3)}$ \\
$T_{\rm eff}$ & 3085 K$^{(2)}$
  & Distance & 98.2$^{+0.15}_{-0.25}$ pc$^{(4)}$ \\
$L_*$ & $0.014\,L_\odot^{(2)}$
  & v$_{\rm LSRK}$ & $5.83$ km\,s$^{-1}$\textsuperscript{(4)} \\
$M_*$ & $0.11\,M_\odot^{(2)}$
  & PA & $111.2^\circ$ \\
$R_*$ & $0.42\,R_\odot^{(2)}$
  & $i$ & $26.3^\circ$ \\
\hline
\end{tabular}
\tablefoot{PA and $i$ denote the disk position angle and
inclination measured in this work.}
\tablebib{
(1) \citet{Mamajek..1999};
(2) \citet{Hashimoto..2025};
(3) \citet{Roccatagliata..2024};
(4) \citet{Gaia..DR3}.
}
\end{table}

\subsection{Contrast curves of NACO and MUSE data in EK Cha}

After applying Spectral Differential Imaging in the MUSE datacube around the H$\alpha$ and H$\beta$ lines, as described in Appendix \ref{App: MUSE}, we reach 5$\sigma$ contrasts of $\Delta\mathrm{mag}(\mathrm{H}\alpha) \approx 7.7$ at the separation of EK Cha b (461 mas),
comparable to the depths reached by similar analyses of MUSE datacubes in the literature
\citep{Xie..2020}. We list the 3$\sigma$ upper limits on the line fluxes in Table \ref{Tab: EK Cha b} based on contemporaneous measurements of the stellar Balmer lines emission using the same MUSE datacube. No statistically significant emission ($>5\sigma$) is retrieved at EK Cha b's location. Nevertheless, accretion variability in protoplanet companions is not unexpected; for example, it has been observed in PDS 70 b and PDS 70 c \citep{Zhou..2025, Laird..2025}, so EK Cha b could still be actively accreting. In addition, dust extinction from the circumplanetary material itself ($A_V=5.0^{+1.0}_{-0.9}$ mag) complicates the detection of lines in the visible such as H$\alpha$ by increasing the contrasts required to reach a constraining threshold on the emission.

     \begin{table}
 \caption[]{\label{Tab: EK Cha b}Emission constraints on EK Cha b.}
\begin{tabular}{lc}
   \hline \hline
  Line/Band &
  Constraint \\ 
   \hline
H$\beta$  & $<$ 5.4$\times$10$^{-9}$   L$_{\odot}$ \\
F606W  & $<$1.05$\times$ 10$^{-19}$ ergs s$^{-1}$ cm$^{-2}$ \AA$^{-1}$  \\
H$\alpha$  & $<$ 2.5$\times$10$^{-9}$  L$_{\odot}$ \\
F814W    &  2.94$\pm 0.01\ \times10^{-18}$ ergs s$^{-1}$ cm$^{-2}$ \AA$^{-1}$\\
\textit{H}-band  & $<$ 9.3$\times$ 10$^{-17}$ ergs s$^{-1}$ cm$^{-2}$ \AA$^{-1}$ \\
\textit{K$_s$}-band  & $<$1.34$\times$ 10$^{-16}$ ergs s$^{-1}$ cm$^{-2}$ \AA$^{-1}$  \\
1.3 mm  & 277$\pm$ 54$\mu$Jy \\
CO J=2--1 & 79.1$\pm$ 14.9 $\mu$Jy  km s$^{-1}$\\
\hline
\end{tabular}
\end{table}

\subsection{EK Cha's properties from the ALMA visibilities fitting}

Despite the low spatial resolution of the ALMA observations, we were able to retrieve the geometrical properties and the radial intensity profile of the disk with the \texttt{galario} and \texttt{frank} packages \citep{galario, Jennings}. In EK Cha, we found the presence of an inner gap (see Figure \ref{Fig: frank}). \citet{Bouwman..2010} suggested a gap-carving planet at the same distance when they modeled EK Cha's infrared SED.  The deep gap found by the fitting of the visibilities is consistent with the disk SED.


\section{Discussion} \label{Sec: Discussion}

\subsection{EK Cha b within the current understanding of planet formation around M dwarfs}

So far, despite their high occurrence rate in exoplanet demographics \citep{Yiang..2020}, \bluebf{and current detections in debris disks around M dwarfs, TWA 7 \citep{Lafreniere+..2007} and RX J0534.0-0221 \citep{Ferrer_chavez}}, M stars remain largely unexplored in protoplanet searches. Thus, EK Cha probes a regime with strong potential for planet formation studies. Compared to wide-separation ($R>$100 au) PMCs, and embedded CPDs such as PDS 70 c \citep{Benisty..21}, EK Cha b appears to belong to a slightly different category. It is beyond the millimeter continuum disk but still embedded in the gaseous emission. Despite the infrared detections of PMCs, ALMA observations often fail to find them in millimeter emission, suggesting that their disks are evolved, compact, or pebble-depleted \citep{Wu..yalin..2020, Facchini..2026}. Conversely, EK Cha b hosts a bright millimeter disk detected with ALMA, broad CO emission, and a detection in a wide band filter centered at $\sim$0.82 $\mu$m.  Relative to the PDS 70 and WISPIT 2 systems, EK Cha is more representative of a typical planet-hosting star, while keeping PMCs's advantage of a clear distinction between the circumstellar disk and the CPD. Thus, EK Cha b probes a different regime of planet formation not explored by the ones mentioned above.

\begin{figure}[h]
    \includegraphics[width=0.45\textwidth]{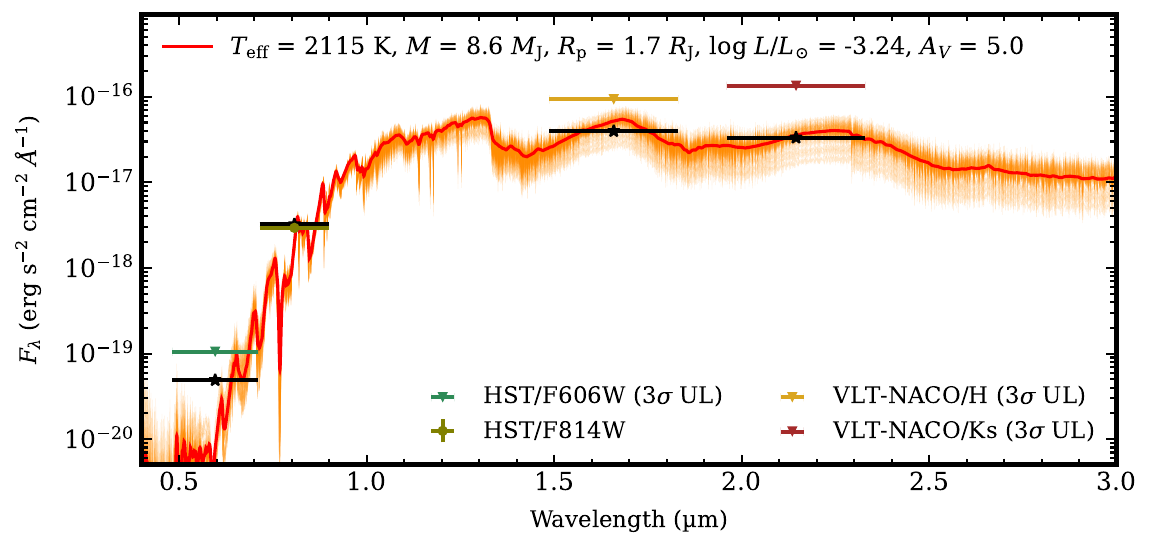} 
\caption{MCMC fitting of the SED with the available constraints using \texttt{species} \citep{Stolker..20} with the \texttt{sonora-diamondback} evolutionary models \citep{Morley..2024}, including the \texttt{Red} extension  \citep{Davis..25}. Top panels shows the filters transmission curves, and the bottom panels the upper limits and the synthetic photometric points for each filter of the best fit in black.}
\label{Fig SED}
\end{figure}

\subsection{The planetary-mass nature of EK Cha b}

 Previous imaging surveys discarded a stellar or brown dwarf companion to EK Cha \citep{Kohler..2002, Brandeker..2006}. As we show in Figure \ref{Fig SED}, when we combine all our available photometric constraints, our SED fitting with atmospheric evolutionary models shows, with a 80\% confidence in \texttt{sonora-diamondback} models, that EK Cha b's mass is below the deuterium burning mass limit, 13 M$_{\rm Jup}$ \citep{Spiegel..2011}. We describe assumed priors and fitting setup in Appendix \ref{App: SED}. Given that no disk emission component was included in the fitting, this mass upper limit of the companion may in fact be lower. A disk emission component would push the near-infrared upper limits for the planetary contribution and the colors as well, leading the fit towards lower effective temperatures and planetary mass solutions. \bluebf{A second epoch, after 11 yr, would allow to confirm the companion given the $>$0.44\arcsec \ displacement of EK Cha due to its proper motion.}
 
 Given that the foreground extinction towards EK Cha is almost negligible \citep{Lyo..06}, we associate the retrieved extinction value with the extinction produced by the CPD itself. Additional photometric points would enable a more thorough description of the CPD's physical conditions and dust properties. This would allow a direct comparison between the primary disk and the CPD around the companion can be performed.

EK Cha b is not only a relevant detection as an intermediate object between typically fully embedded protoplanets and wide-orbit planetary-mass companions, but it also showcases the potential of evolved disks around late-type stars, with low foreground and circumstellar disk extinction, exposing the planet-forming regions, even in the last stages of their evolution. 

\section{Summary}

We reported multi-tracer detections of EK Cha b in spatially and spectrally resolved CO emission, 1.3 mm continuum, and F814W photometry. Its kinematics and SED are consistent with a planetary-mass companion surrounded by a circumplanetary disk, although deeper observations  and/or multiple epochs are required to fully understand and confirm its nature. Photometric data will set stronger constraints on mass and temperature, while ALMA can inform about the CPD's properties. EK Cha b highlights that evolved disks around low-mass stars are promising targets for future protoplanet searches and CPD studies.

\begin{acknowledgements}
    Based on observations collected at the European Organisation for Astronomical Research in the Southern Hemisphere under ESO programme(s) 70.C-0286(A), 90.C-0649(A), and 108.C-0151(B).
    This paper makes use of the following ALMA data: ADS/JAO.ALMA\#2017.1.01419.S. and ADS/JAO.ALMA\#2021.1.01205.S. ALMA is a partnership of ESO (representing its member states), NSF (USA) and NINS (Japan), together with NRC (Canada), NSTC and ASIAA (Taiwan), and KASI (Republic of Korea), in cooperation with the Republic of Chile. The Joint ALMA Observatory is operated by ESO, AUI/NRAO and NAOJ.
    This research is based on observations made with the NASA/ESA Hubble Space Telescope obtained from the Space Telescope Science Institute, which is operated by the Association of Universities for Research in Astronomy, Inc., under NASA contract NAS5–26555. These observations are associated with program(s) 14212.
      FA, SC and SF are funded by the European Union (ERC, UNVEIL, 101076613). Views and opinions expressed are however those of the author(s) only and do not necessarily reflect those of the European Union or the European Research Council. Neither the European Union nor the granting authority can be held responsible for them. SF and SC acknowledge financial contribution from PRIN-MUR 2022YP5ACE.
\end{acknowledgements}


\bibliography{ref}{}
\bibliographystyle{aa}

\begin{appendix}
    \section{Observational Data} \label{App: Data}
    \subsection{HST/ACS} \label{App: HST}

    The HST-ACS observations are part of the program 14212, and they were taken on August 29th of 2017 with 77 seconds of exposure time for the F814W filter and 27 seconds for the F606W filter. The raw data files were run through the standard STScI ACS pipeline including the imaging drizzling step \citep[\texttt{AstroDrizzle}, ][]{Gonzaga..2012}. Background emission was estimated by taking  the median of an annulus centered in EK Cha with an inner radius of 2 arcsec and then subtracted. For the companion, we used a 3 pixels radius aperture photometry and applied a correction by dividing by the enclosed flux in ACS-F814W observations with 3 pixel apertures, which is 0.764 \citep{Sirianni..2005}. The aperture for the companion was smaller to avoid flux contribution from EK Cha's emission. The uncertainty estimations were done by first taking random non-overlapping apertures in the field-of-view and performing the photometry on them with the same aperture used for the companion (3 pixels). Then, from the resulting distribution of photometric values, we discard values above the 90\% and below the 10\% quantiles of the distribution to avoid outliers, such as random apertures falling on others stars in the field of view. From the resulting distribution we calculate the standard deviation as a proxy of the noise. The reported value in Table \ref{Tab: EK Cha b} for the F606W filter corresponds to three times (3$\sigma$) the value retrieved from this method, while the uncertainty in the F814W measured flux for the companion's detection is the retrieved value through this method (1$\sigma$).

    To check that the feature did not correspond to  a Point-Spread Function (PSF) feature, we performed a PSF subtraction taking into account the stellar emission. The PSF model was done using the \texttt{TinyTim} package \citep{TinyTim}, which was developed to produce synthetic PSFs for HST. The spatial resolution reached by HST, represented through the Full Width  at Half Maximum (FWHM) for the F814W filter is $\approx$ 84 mas. We showed the respective PSF model and the residual image once the star was subtracted in Fig. \ref{Fig: PSF_SUB}.

    \begin{figure}[h]
    \includegraphics[width=0.4\textwidth]{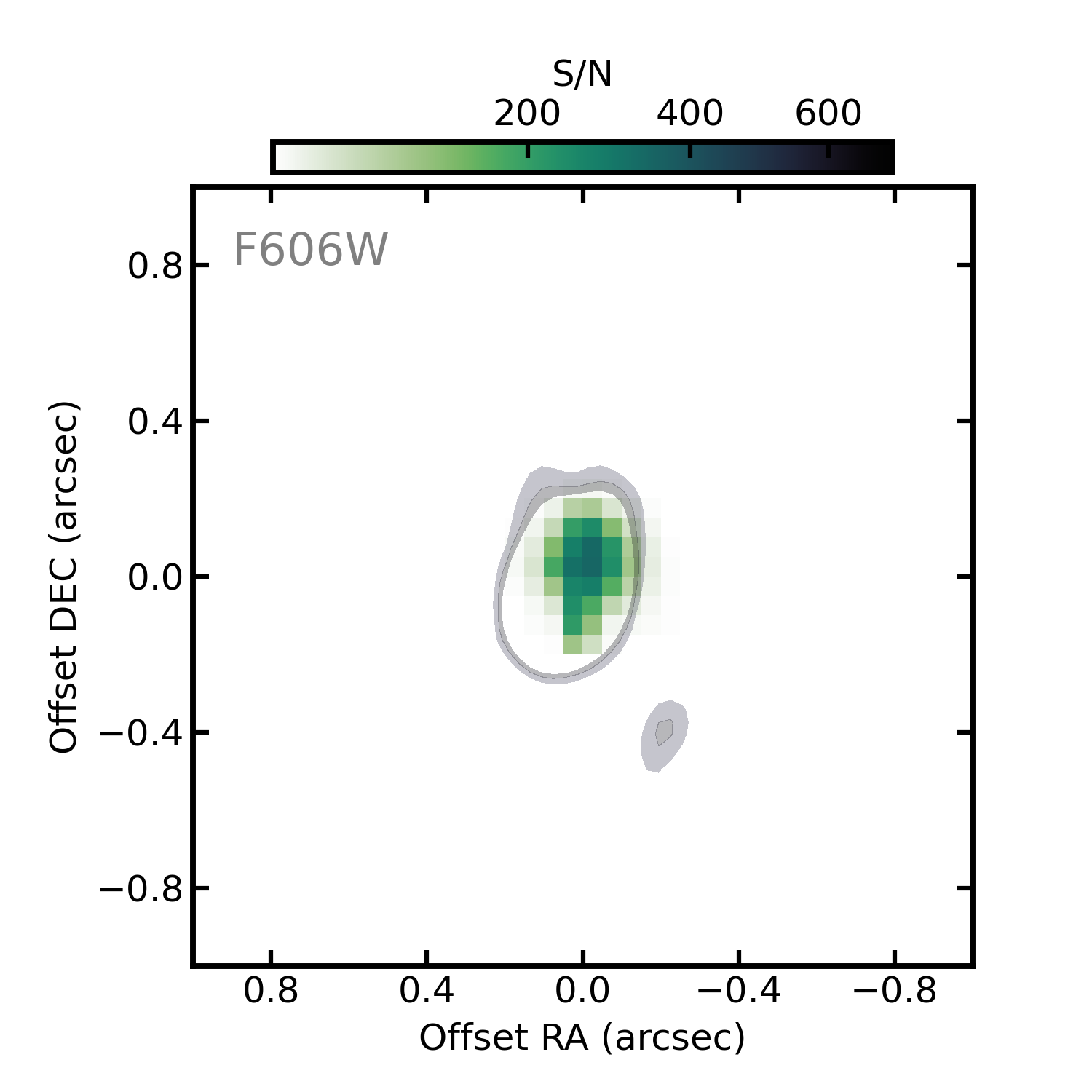} 
\caption{HST/ACS image of EK CHa with the F606W filter. Overlaid are the contours from the 1.3 mm continuum observations with ALMA.}
\label{Fig: HST-V}
\end{figure} 
    
    Even though a second epoch observation with multiple dithering  would be required to robustly discard the chances of cosmic-ray hits, the spatial coincidence with the ALMA detection makes a chance of a random cosmic-ray hit unlikely. The F606W observation illustrated in Fig. \ref{Fig: HST-V} does not show any counterpart to the emission detected in the infrared and millimeter continuum with high significance. The far-red optical color upper-limit, F606W - F814W ($>$5.82), together with the negligible extinction towards EK Cha provides strong evidence for the nature of the companion.

    We derived HST's astrometric precision by cross-matching sources in the HST/ACS-F814W field of view against the Gaia catalogue, resulting in four common objects. The registration uncertainty was then estimated from the standard deviation of the corrections for each individual star under a leave-one-out cross-validation, giving 20 mas.

    The features in the HST/ACS-F814W and 1.3 mm ALMA images are not perfectly spatially coincident, showing a separation of $\sim$80 mas. In HST has a separation of 451 mas  and a PA of 190 \degr, and in ALMA 461 mas with a PA of 200 \degr. Combining the ALMA astrometric precision of the EK Cha b emission (42 mas) in quadrature with our HST registration uncertainty (20 mas) gives a combined inter-observatory precision of 47 mas. We consider this value a lower limit  since some additional sources of error were not fully accounted for; however, these are not expected to change the final relative astrometric precision significantly. Given this combined uncertainty, a 80 mas separation corresponds to a 1.7 $\sigma$ offset, therefore it is still consistent with the two features being co-located.

    \begin{figure}[h]
    \includegraphics[width=0.49\textwidth]{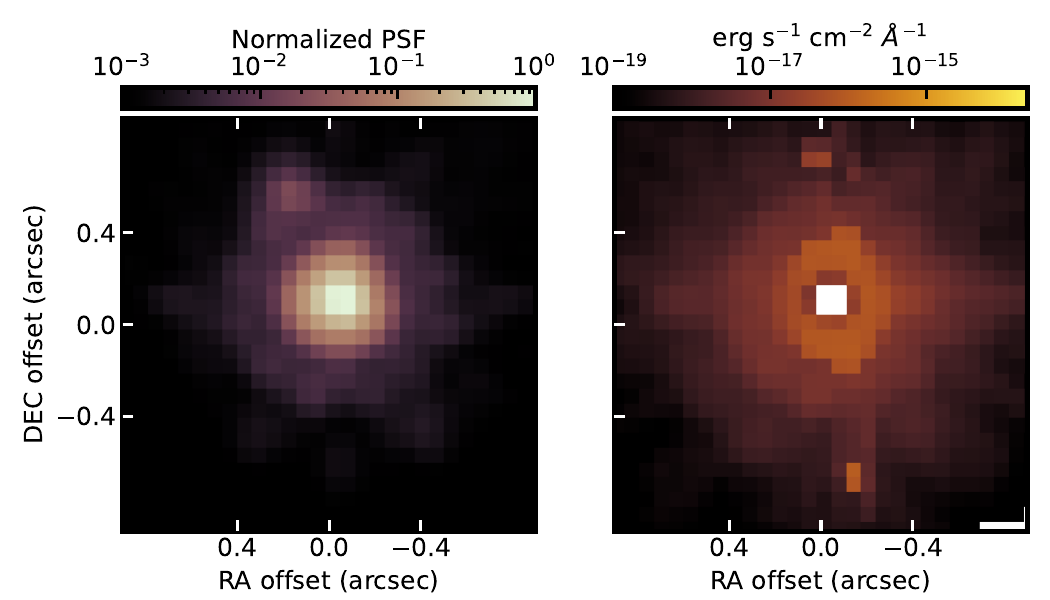} 
\caption{\textbf{Left:} PSF model of the HST F814W observations of a M-dwarf point source produced with \texttt{TinyTim}. \textbf{Right:} Residual HST/ACS-F814W image of the EK Cha system after subtracting the stellar emission.}
\label{Fig: PSF_SUB}
\end{figure} 

    \subsection{VLT-NACO} \label{App: VLT}

    The VLT-NACO \textit{H}-band observations were taken as part of the program \texttt{70.C-0286(A)} with 52 min of integration time. The results of those observations and the contrast curve were reported by \citet{Brandeker..2006}. We used their 5$\sigma$ sensitivity value for EK Cha at the separation of EK Cha b as an upper limit for our SED analysis. 

\begin{figure}[h!]
    \begin{center}
    \includegraphics[width=0.49\linewidth]{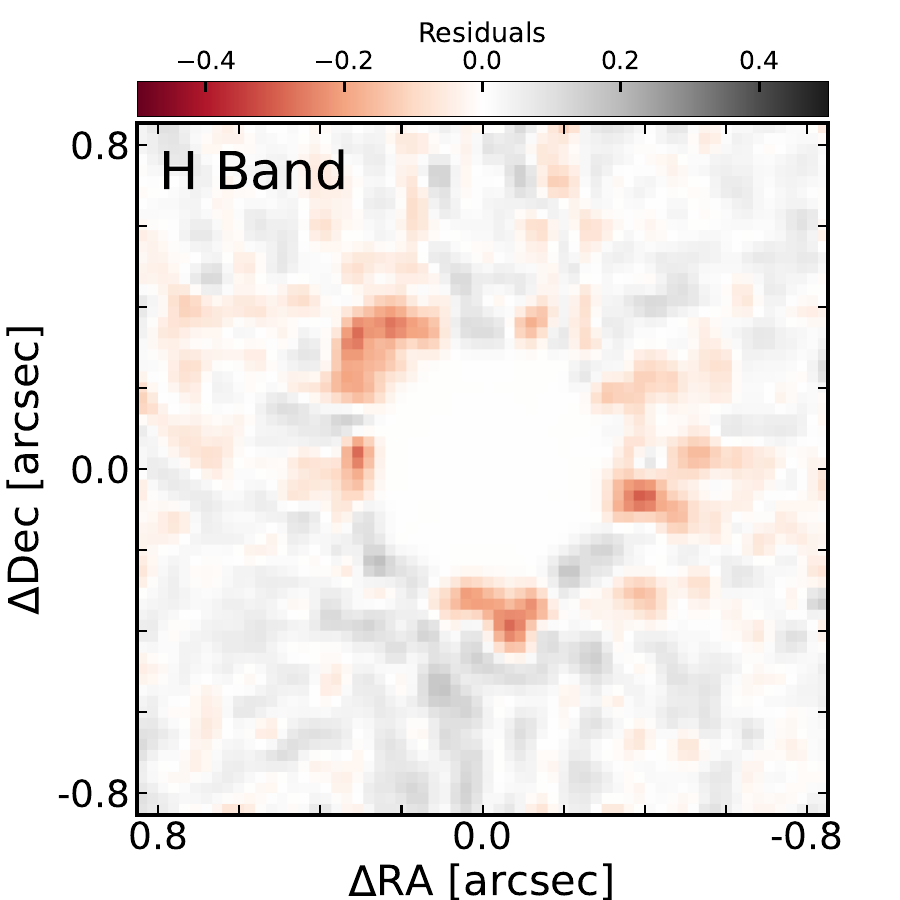}
    \includegraphics[width=0.49\linewidth]{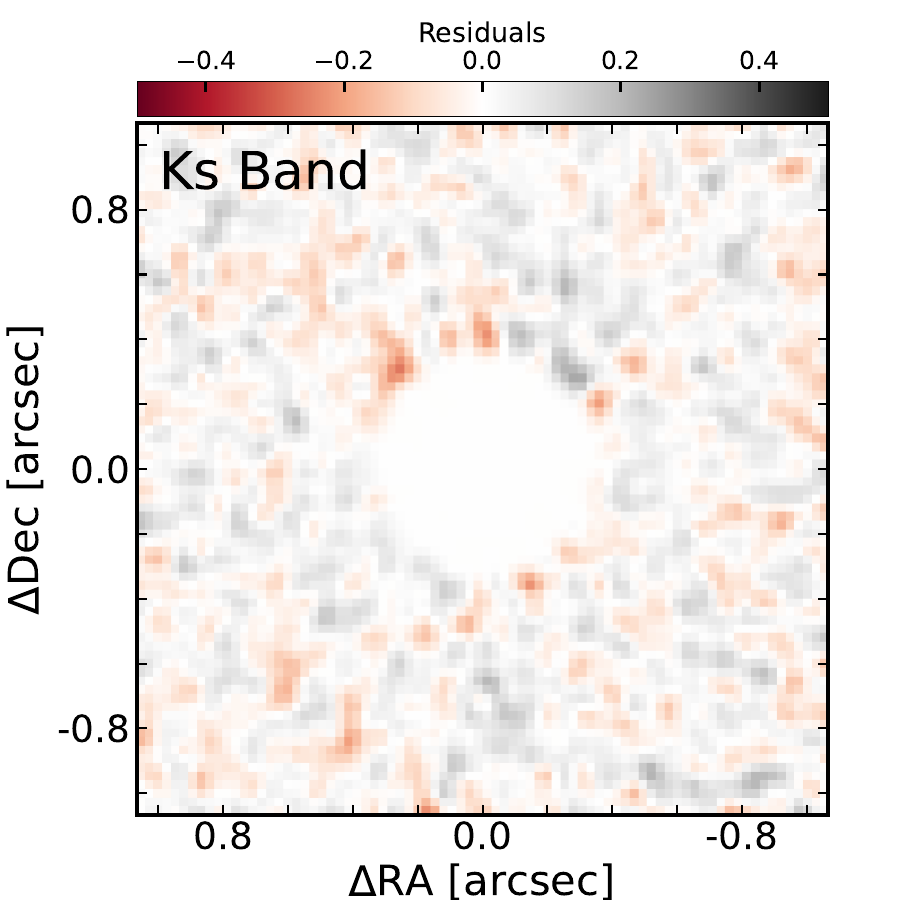}
    \end{center}
    \caption{Residuals of the \textit{H} band (left) and \textit{K$_s$} (right) emission after performing high-contrast imaging on the VLT-NACO data of EK Cha.}
    \label{fig: NACO-HCI}
\end{figure}

    Regarding the \textit{K$_s$}-band observations, EK Cha was also observed with VLT/NACO in program \texttt{0.90.C-0649(A)} with a total integration time of 56 min. We calibrated the raw dataset using the \texttt{EsoRex} pipeline. Once we obtained the science-ready frames, we applied a standard Angular Differential Imaging \citep[ADI, ][]{Marois..2006} reduction with the \texttt{VIP} package for high-contrast imaging \citep{VIP1, Vip2}. The pipeline involved recentering frames, discarding frames with poor quality and flagged by the pipeline. Then, we performed the PSF characterization followed by the ADI subtraction through an annular PCA algorithm. We explored the number of PCA components that would maximize the contrast at the given separation  with throughput correction using the injection  of fake-companions. The optimal value of PCA components was three. We show the resulting images after PCA subtraction in Fig. \ref{fig: NACO-HCI}. Finally, the contrast curve was calculated with the same annular PCA from four position angles. We also took into account the Student-t correction in the inner regions of the field next to the star fron contrast calculations \citep{Mawet..2014}. The final value of the contrasts reached are listed in Table \ref{Tab: EK Cha b}.

\subsection{MUSE}\label{App: MUSE}
We used a VLT/MUSE-NFM dataset from the ESO science-ready data archive. The dataset corresponds to program Id \texttt{108.22} and it was observed in August 2022 with a total exposure time of 150 min. For each Balmer line (H$\alpha$, H$\beta$) we built a narrow-band line image by averaging the three channels covering the line core ($\sim$3.75\,\AA{} extraction window) and a continuum
reference by median-combining the $\sim$25 line-free channels in the wings, symmetrically placed at $7\,\AA \le |\Delta\lambda| \le 25\,\ \AA$ from the line centre with a guard interval to avoid contamination by the line
wings. Before subtraction, each wing channel was geometrically rescaled about the stellar centroid by $\lambda_\mathrm{line}/\lambda_\mathrm{wing}$ so that
its diffraction-limited speckle pattern registers with that of the line image \citep[e.g.][]{Sparks..2002}.

\begin{figure}
    \centering
    \includegraphics[width=0.49\linewidth]{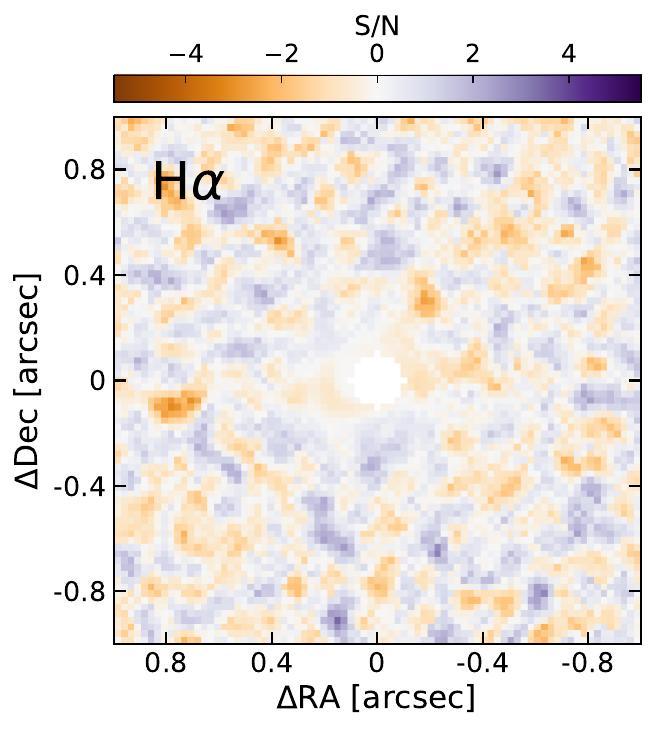}
    \includegraphics[width=0.49\linewidth]{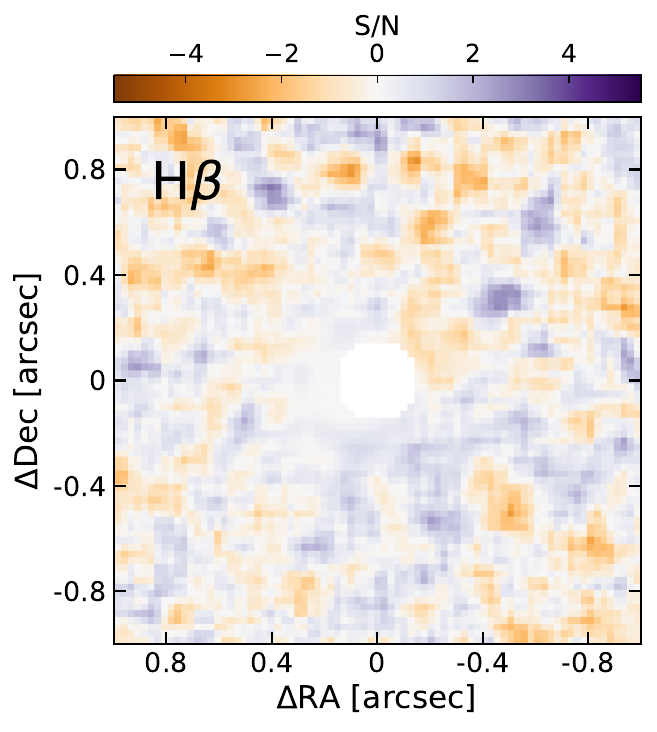}
    \caption{Residuals of the H$\alpha$ (left) and H$\beta$ (right) after performing high-contrast imaging on the VLT-MUSE data of EK Cha.}
    \label{fig:: MUSE-HCI}
\end{figure}

\begin{figure}[h]
    \includegraphics[width=0.5\textwidth]{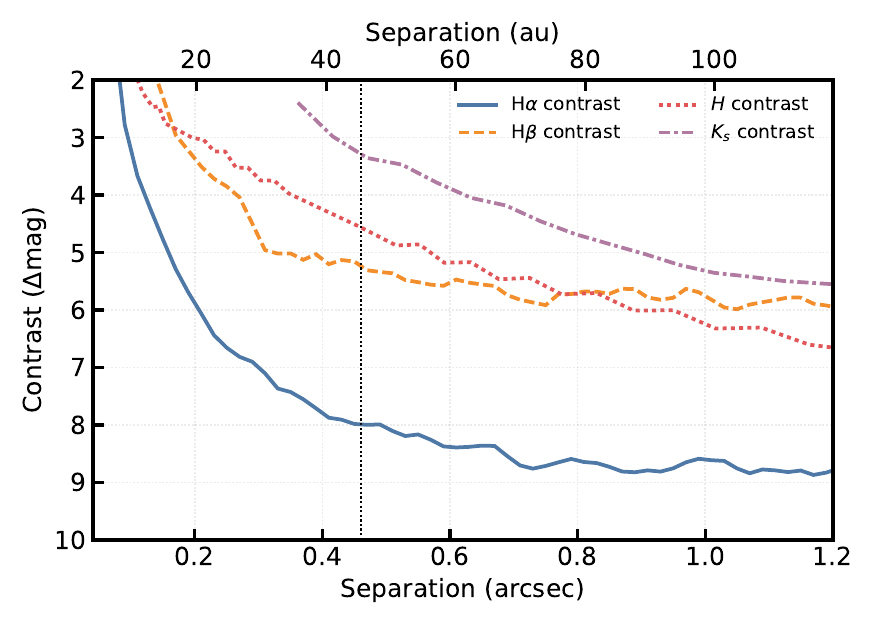} 
\caption{5$\sigma$ contrast curves for EK Cha combining literature values  for H-band \citep{Brandeker..2006} and K-band \citep{Kohler..2002} together with retrieved values of archival data for Balmer lines from VLT-MUSE and K$_s$ data from VLT-NACO. The vertical dotted lines is the separation of EK Cha b.}
\label{Fig:contrast}
\end{figure}

PSF subtraction was performed via a Karhunen--Lo\`{e}ve projection
\citep[KLIP;][]{Soummer..2012}: the line image is modelled as a linear combination of the first $N=5$ principal components of the rescaled wing cube, with the fit restricted to an annulus $2\,\mathrm{FWHM} < r < 1.5\arcsec$ to keep the stellar PSF core out of the basis fit. The resulting residuals are shown in Fig. \ref{fig:: MUSE-HCI}. The 5$\sigma$ point-source contrast was then derived by matched-filtering the residual with a top-hat
aperture of radius $\mathrm{FWHM}/2$, measuring the noise in concentric annuli with a robust MAD estimator, and applying the \citet{Mawet..2014} Student-$t$ correction for the small number of independent resolution elements at small separations. Algorithmic throughput was calibrated by injection--recovery of artificial point sources at six position angles per
separation; the recovered-to-injected ratio stayed in the range 0.95--1.00 across the explored separations and was applied as a multiplicative correction to the contrast curve. As a cross-check we also ran the single-scalar prescription $I_\mathrm{line} - \alpha\,I_\mathrm{cont}$ \citep[][with $\alpha$ fit by least squares in an off-star annulus]{Xie..2020}, which agrees with the KLIP result to within 20\% across the explored range. The resulting 5$\sigma$ $\Delta$mag upper limits
are shown in Figure~\ref{Fig:contrast}. At separation of EK\,Cha\,b we only recovered emission at 1$\sigma$ level in both Balmer lines, which is not statistically significant to claim a detection.

We also retrieve the H$\alpha$ and H$\beta$ stellar line fluxes from the datacube as they are the basis for the contrast values that we report. The line fluxes were retrieved by using an aperture photometry with a radius equal to 2$\times$ the FWHM at the respective wavelength and then subtract the continuum emission estimated from the line-free adjacent channels. The same channel range that was used for our HCI approach was used for continuum subtractions. Then, we visually inspected the line profiles to check for significant asymmetries. As we did not find significant asymmetries in the line profile, we measured the line flux by fitting a Gaussian profile in the spectral axis and integrating over the fitted profile.

    \subsection{ALMA} \label{App:ALMA}

    The ALMA datasets correspond to two separate programs that observed EK Cha at 1.3 mm (Band 6) and 0.8 mm (Band 7). The Band 6 data correspond to the program \texttt{2017.1.01419.S} and includes the detection of the CO J=2--1, $^{13}$CO J=2--1 and C$^{18}$O J=2--1 rotational lines with a total of 20.7 min on source at medium spatial resolution. EK Cha was also observed in Band 7 as part of the \texttt{2021.1.01205.S} ALMA program with  8.6 min of time on source in a compact configuration. The spectral setup covered the CO J=3--2, HCO$^{+}$ J=4--3  and HCN J=4--3 lines, which were also detected. The rest frequencies and native spectral resolution for each line are listed in Table \ref{Tab: ALMA}.

    \begin{table}
 \caption[]{\label{Tab: ALMA}Emission lines detected in ALMA data.}
\begin{tabular}{lccccc}
   \hline \hline
  Line &
  Rest Frequency &
  Spectral Resolution\\ 
   &
  (GHz) &
  (kHz)   \\ 
   \hline
 \noalign{\smallskip}
\multicolumn{3}{c}{Band 6} \\
\noalign{\smallskip}
\hline
CO J=2--1   & 230.5380  & 122.070 \\
$^{13}$CO J=2--1    &  220.3987 & 488.281\\
C$^{18}$O J=2--1   & 219.5604 & 488.281 \\
\hline
\noalign{\smallskip}
\multicolumn{3}{c}{Band 7} \\
\noalign{\smallskip}
\hline
CO J=3--2   & 345.7960 & 122.070 \\
HCN J=4--3  &  354.5055 & 488.281   \\
HCO$^{+}$ J=4--3 & 356.7342  & 244.141 \\
\hline
\end{tabular}
\end{table}

\begin{figure}
    \includegraphics[width=0.485\textwidth]{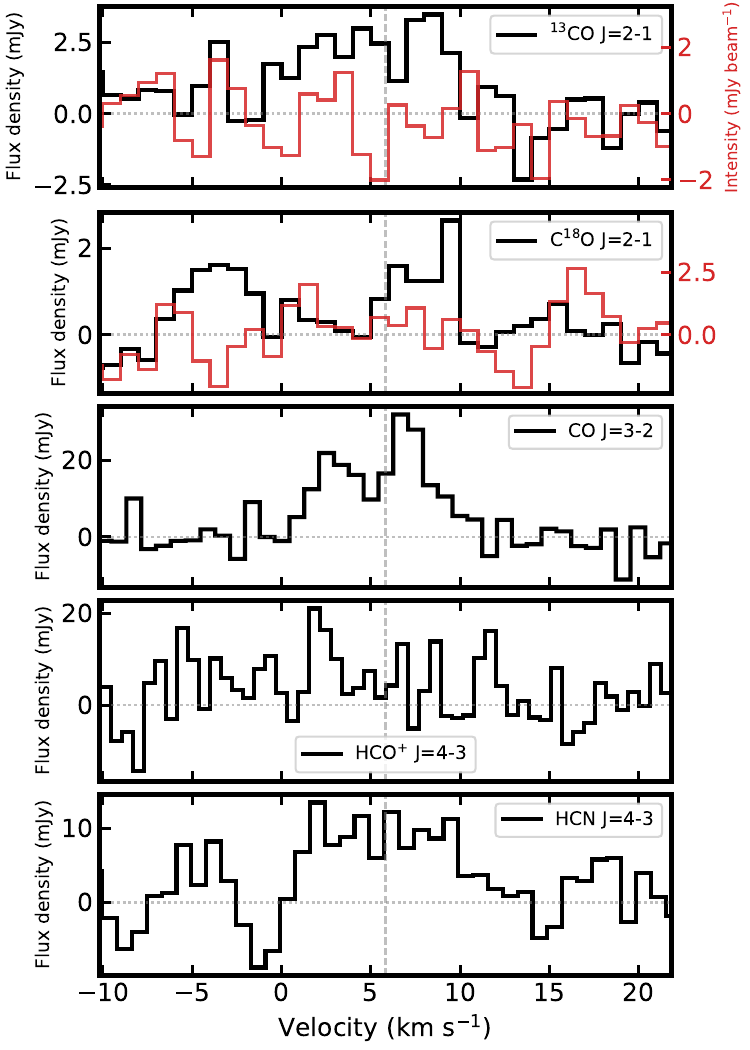} 
\caption{Ek Cha's spectra for the detected lines in the ALMA datasets. Vertical dashed line is the systemic velocity of EK Cha.}
\label{Fig :spec_alma}
\end{figure}   

    \begin{table}
 \caption[]{\label{Tab: Beam}EK Cha's line/continuum fluxes observed with ALMA.}
\begin{tabular}{lccccc}
   \hline
   \hline
 \noalign{\smallskip}
 \multicolumn{3}{c}{Dust Continuum} \\
\noalign{\smallskip}
\hline
  Continuum Band &
  Flux &
  Beam \\ 
   &
  (mJy) &
    (B$_{\rm min}$ $\times$ B$_{\rm maj}$, PA)\\ 
   \hline
Band 6 (1.3 mm)   & 5.36 $\pm$ 0.09 & (0\farcs15$\times$0\farcs23, 107\degr ) \\
Band 7 (0.8 mm)  &  12.11 $\pm$ 0.15 & (0\farcs29$\times$0\farcs52, 128\degr )   \\
\hline

\noalign{\smallskip}
\multicolumn{3}{c}{Molecular Line Emission} \\
\noalign{\smallskip}
   \hline 
  Line &
  Flux &
  Beam \\ 
   &
  (mJy km s$^{-1}$) &
  (B$_{\rm min}$ $\times$ B$_{\rm maj}$, PA)   \\ 
   \hline
 \noalign{\smallskip}
CO J=2--1   & 426$\pm$60 & (0\farcs 15$\times$0\farcs23, 107\degr ) \\
CO J=3--2   & 474$\pm$73 & (0\farcs3$\times$0\farcs57, 128\degr ) \\
$^{13}$CO J=2--1    &  67$\pm$32  & (0\farcs15$\times$0\farcs23, 108\degr )\\
C$^{18}$O J=2--1  & 40$\pm$23 & (0\farcs21$\times$0\farcs31, 98\degr ) \\
HCO$^{+}$ J=4--3   & 210$\pm$118  & (0\farcs29$\times$0\farcs55, 128\degr ) \\
HCN J=4--3   & 268$\pm$48 & (0\farcs3$\times$0\farcs56, 128\degr ) \\
\hline
\end{tabular}
\end{table}

\begin{figure}[h]
    \includegraphics[width=0.5\textwidth]{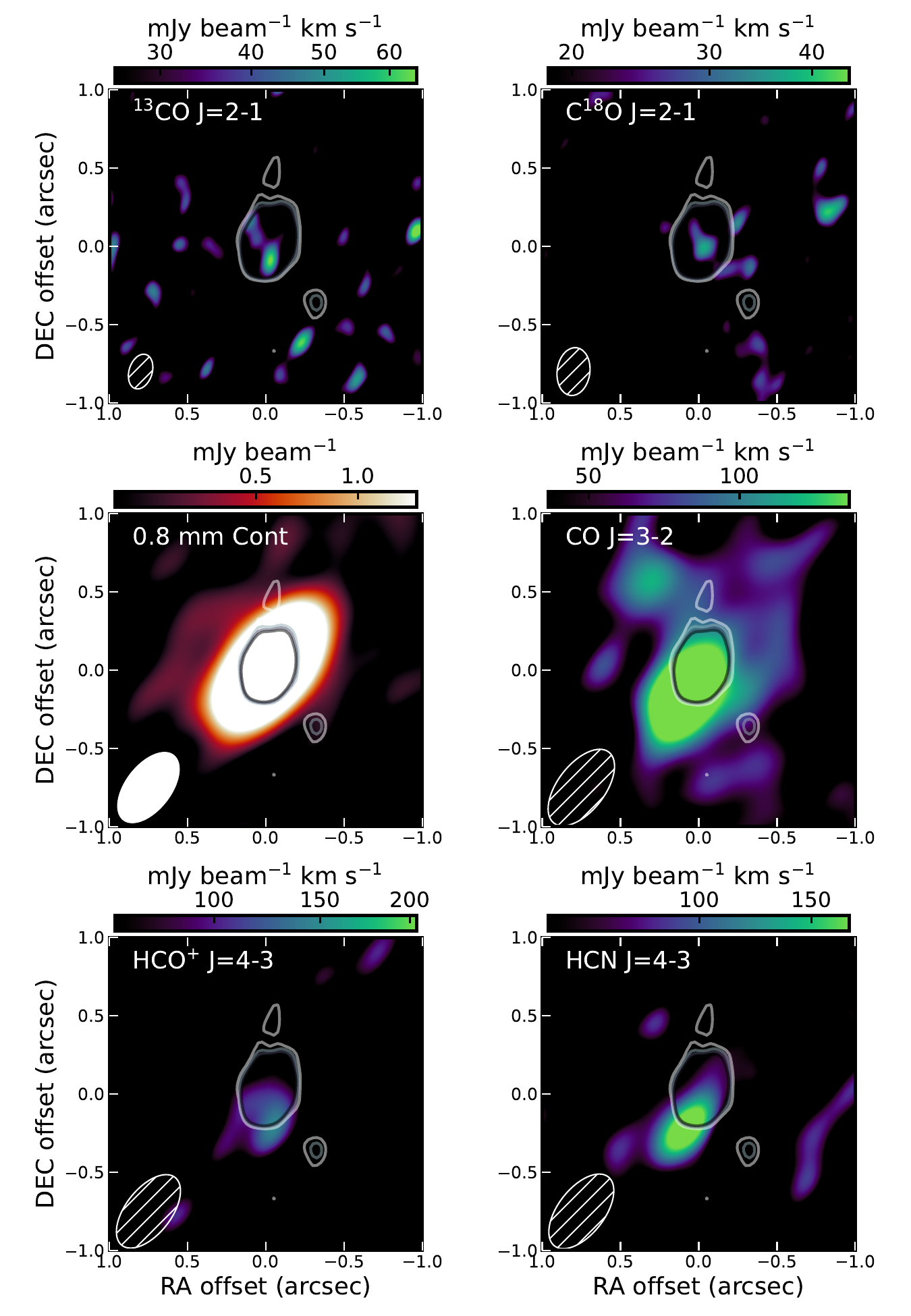} 
\caption{Integrated flux density of detected lines in the ALMA dataset and dust continuum emission at 0.8 mm. For references, the contour from the ALMA 1.3 mm are overlaid and the respective beams are shown on the bottom right of each panel.}
\label{Fig: moms}
\end{figure}

    For both datasets, we performed self-calibration to improve the data quality. For the 1.3 mm data we applied a first round with an \texttt{inf} solution interval and then a solution interval of \texttt{180 s}.  After the two rounds of self-calibration we reach a continuum noise-level of 32.6 $\mu$Jy beam$^{-1}$. For the 0.8 mm data we only performed one round of self-calibration using the wide spectral windows with an \texttt{inf}  solution interval. The resulting noise-level is 110 $\mu$Jy beam$^{-1}$. The calibration solutions were then applied to the full datasets. The spectral windows showing line features were continuum-subtracted in the \textit{uv}-plane using the \texttt{uvcontsub} routine in CASA. Then, we imaged  each line present on each dataset respectively with \texttt{robust} parameter of 0.5 in \texttt{tclean} and a pixel size  $\sim$1/8 of the beam size. The only exception was CO J=2--1 where we used \texttt{robust}=1.0 to improve the signal-to-noise ratio. We then integrated the emission inside the disk with an elliptical mask defined by the disk's continuum emission in each band. The integrated spectra for the different lines are shown in Figure \ref{Fig :spec_alma}. We also show the respective moment 0 maps and the 0.8 mm continuum in Figure \ref{Fig: moms}. We list all the line emission fluxes and millimeter continuum emission together with their respective beam dimensions in Table \ref{Tab: Beam}. In Band 6 data we can place an upper limit of $\sim$20 au for the CPD's radius since it is unresolved. The uncertainties are retrieved by sampling multiple apertures in emission free regions of the Field of View as described by \citet{Alarcon..2026}. We also include the relevant H$\alpha$ and H$\beta$ line fluxes from the MUSE datacube as they provide the absolute upper limits retrieved by our contrasts.

    \section{SED Fitting} \label{App: SED}

    We perform an SED fitting of the available photometric constraints using the \texttt{species} package \citep{Stolker..20} and the \texttt{dynesty} nested sampler \citep{Dynesty}. \texttt{species} uses a database of atmospheric evolution models from stellar to planetary mass objects to fit for the correspondent observational constraints.  The fitting is performed using the detection from the F814W filter and the upper limits from the broadband filters, i.e., the F606W, the \textit{H} band filter, and the \textit{K$_s$} filter. In the fitting, the free parameters are the effective temperature of the object $T_{\rm eff}$, the logarithm of the surface gravity, and the extinction in the visible $A_V$. For this fitting we used the extinction law profiles from \citet{Gordon..23} and the \texttt{sonora-diamondback} atmospheric evolutionary models \citep{Morley..2024}  including the \texttt{Red} expansion \citep{Davis..2025}. Among the different evolutionary models we opted for the \texttt{sonora-diamondback} as it covers the range from massive young planets to low mass stars and the emission from optical to infrared, which are the values we used to constraint EK Cha b's properties. Since Gaia's measurement of EK Cha's iron abundance is 0.13, we assumed a solar metallicity value for the atmospheric fitting, i.e. \texttt{feh}=0.0, and restrict the cloud coverage factor to \texttt{fsed} $\in$ [1,2] due to the low age of the system. The companion's mass and radius are retrieved from the evolutionary models themselves. The distance, $d$, is fixed at 98.2 pc, which is the distance to EK Cha (see Table \ref{Tab: EKCHa_pars}). 

    We use uniform priors ranges as broad as the range covered by the models overlapping with the ranges $T_{\rm eff} \in$ [500,2800] K, log$(g) \in$ [3.4,4.5] ($g$ in cm s$^{-2}$), and $A_V \in$ [0,15] mag. The fit was performed through a random walk sampling using 512 live points and a multi-ellipsoidal bounding method. We used an \texttt{evidence\_tolerance=0.25} as a convergence criterion. We show the resulting posterior distributions in Figure \ref{Fig: corner}. For this instance, the companion's luminosity, radius and mass are the ones derived from each evaluation of the log$(g)$ and T$_{\rm eff}$ from the evolutionary models. As the figure illustrates, there is a strong constraint on the thermal upper limit imposed by the HST-F606W non-detection, while the lower limits are constrained by the infrared ground-based observations.

    We also show the evolutionary track obtained also for \texttt{sonora-diamondback} models, but using the inferred luminosity from the previous fit. For the evolutionary track we used a two-piece normal distribution centered at 8 Myr with a positive standard deviation of +1 Myr and a negative standard deviation of -2.5 Myr. This assumption is based on the fact that EK Cha b was probably formed after the formation of EK Cha itself rather than before, so it is constrained by EK Cha's age, although it is still unknown when in particular since EK Cha's formation. Thus, we increase the standard deviation on the age prior for EK Cha b. The luminosity priors was assumed a two-piece normal distribution using the values from the posterior in Fig. \ref{Fig: corner} obtained from the SED fitting.

\begin{figure}[h]
    \includegraphics[width=0.5\textwidth]{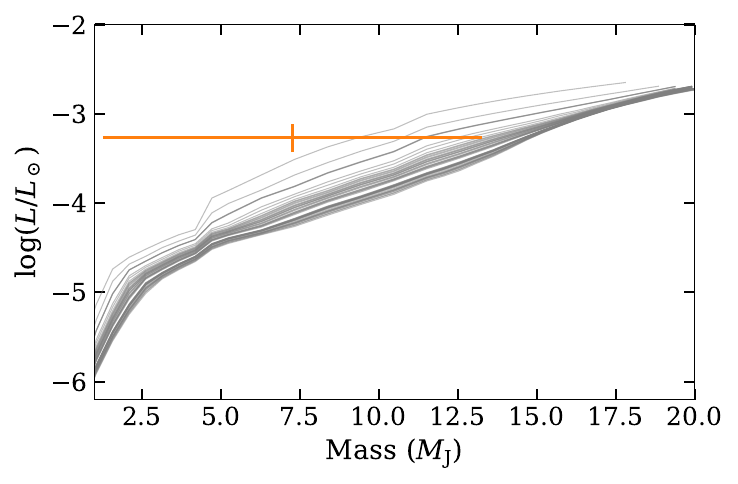} 
\caption{Isochrones of the \texttt{sonora-diamondback} models using the luminosity posteriors as prior, and EK Cha's age as upper limit.}
\label{Fig: evol}
\end{figure}

\begin{figure}[h]
    \includegraphics[width=0.5\textwidth]{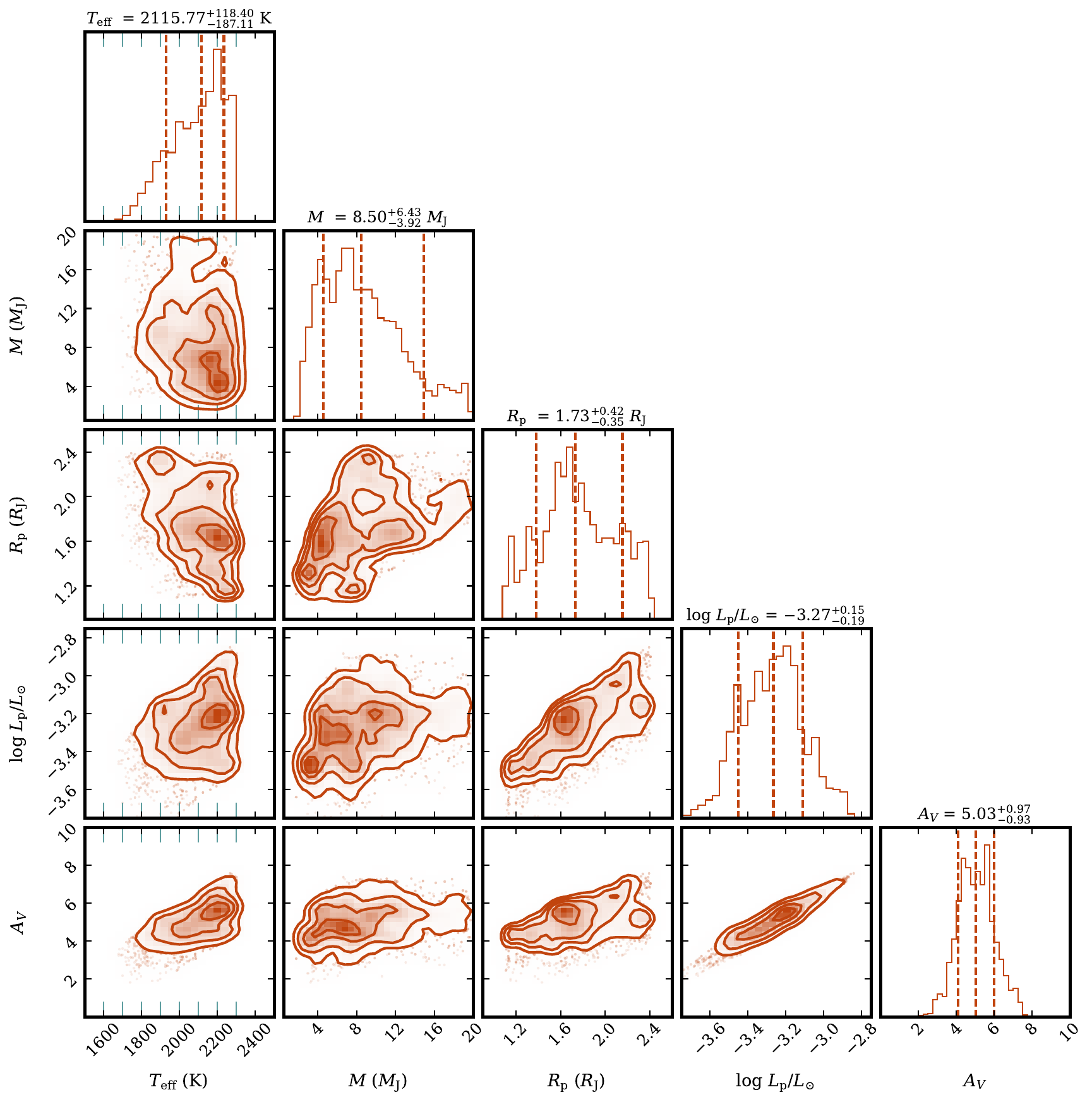} 
\caption{Posterior distributions of the parameters from the performed fit using the \texttt{species} package and the \texttt{sonora-diamondback} models including the \texttt{Red} expansion \citep{Morley..2024, Davis..2025}.}
\label{Fig: corner}
\end{figure}

    We show the evolutionary track obtained from the \texttt{sonora-diamondback} in Fig. \ref{Fig: evol}. Given the initial high luminosity of EK Cha b and the strong constraints from the age, it is possible that the extinction estimations may in fact be overestimated, placing EK Cha b within the same planetary range than the one retrieved by performing the SED fitting. On the contrary, underestimation of the dust extinction would lead to higher companion masses, although that would cause hotter temperatures that would be inconsistent with the available constraints.

\begin{figure}[h]
    \includegraphics[width=0.5\textwidth]{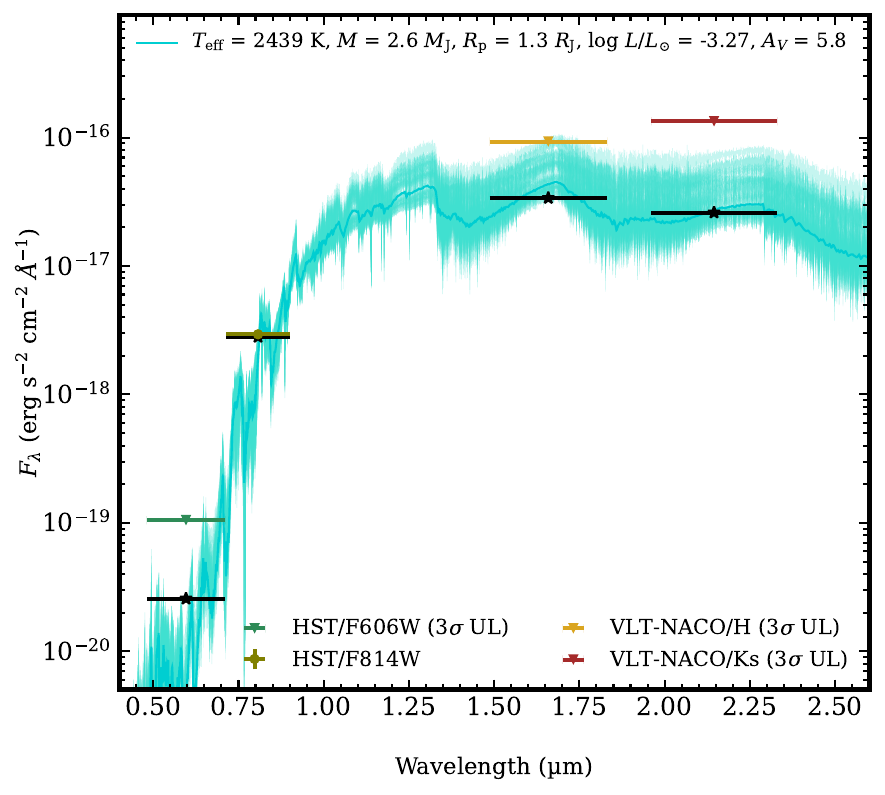} \\
\includegraphics[width=0.5\textwidth]{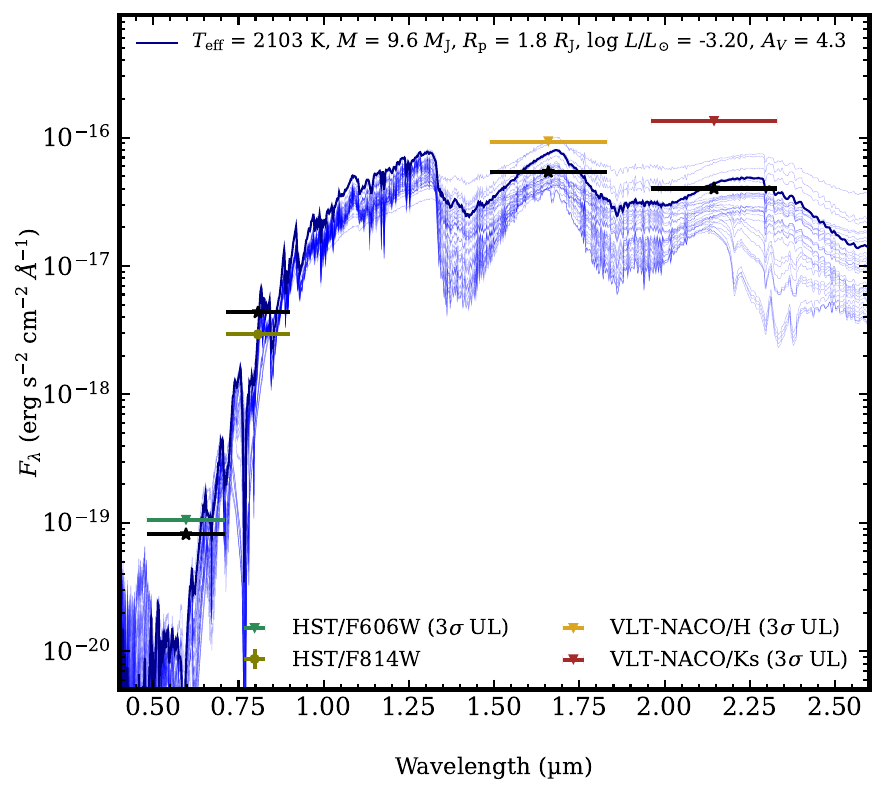}
\caption{Best SED fits from the available observational constraints using the \texttt{species} package for \texttt{BT-settl-cifist} \citep{BT_Settl} in turqouise on the top and \texttt{ATMO} \citep{ATMO} in blue at the bottom.}
\label{Fig: SED_BT_ATMO}
\end{figure}

\begin{figure}[h]
\includegraphics[width=0.5\textwidth]{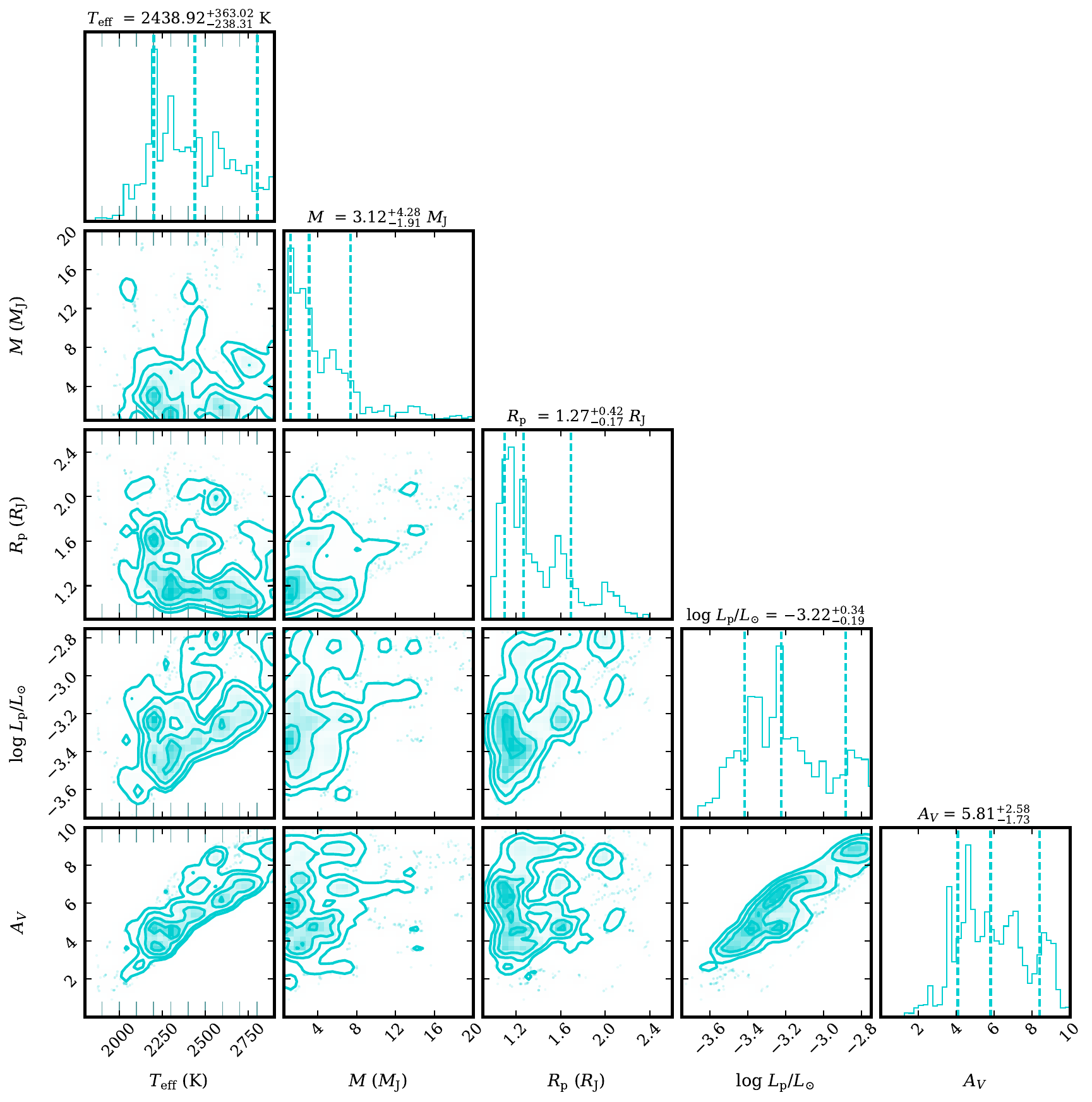} \\
\includegraphics[width=0.5\textwidth]{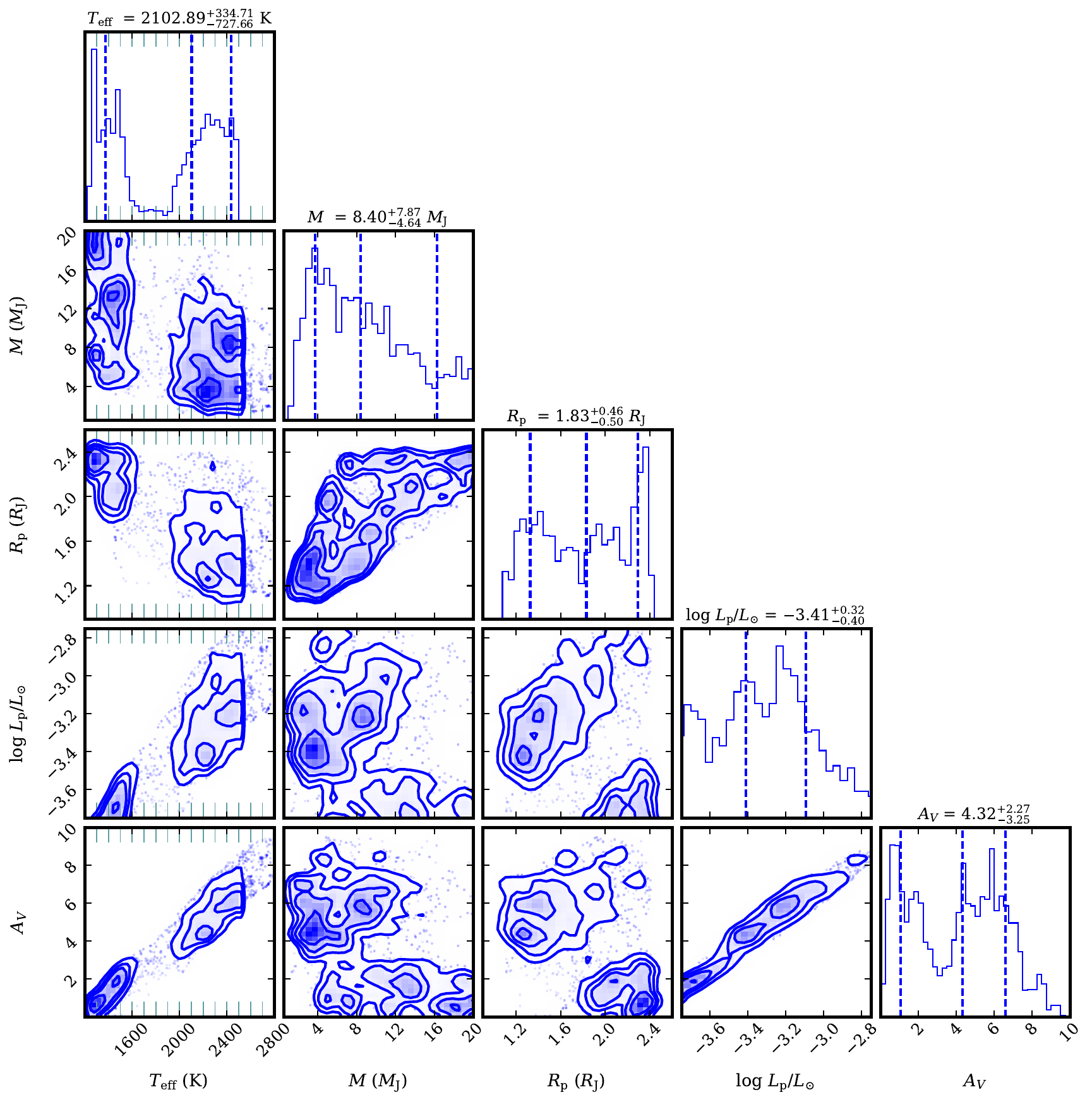}
\caption{Posterior distributions of the parameters from the performed fit using the \texttt{species} package for \texttt{BT-settl-cifist} \citep{BT_Settl} in turqouise on the top and \texttt{ATMO} \citep{ATMO} in blue at the bottom.}
\label{Fig: corner_atmo_bt}
\end{figure}

We also tested with two other atmospheric evolutionary models, which are \texttt{ATMO} \citep{ATMO} and \texttt{BT-settl-cifist} \citep{BT_Settl} that covered a broad range of surface gravities and surface temperatures that may represent  the ones of young planets. We show the respective SED fits and posterior distributions for the two models in Figs. \ref{Fig: SED_BT_ATMO} and \ref{Fig: corner_atmo_bt} respectively, with \texttt{BT-settl-cifist} in turqouise and \texttt{ATMO} in blue. Overall, the three models show roughly consistent results regarding the ranges for the mass of EK Cha b, and a median for the posterior within the planetary-mass range. More significant changes are associated to the extinction and effective temperature. Remarkably, \texttt{BT-settl-cifist} finds lower mass solution since it probes a range with lower surface gravity values that represent hot and large young giant planets. More photometric points and more sensitive constraints can help improve the fit, setting more stringent constraints on the limits of the parameters of EK Cha b.

    \section{Fitting of ALMA visibilities} \label{App: Frank}

\begin{figure}[h]
    \includegraphics[width=0.5\textwidth]{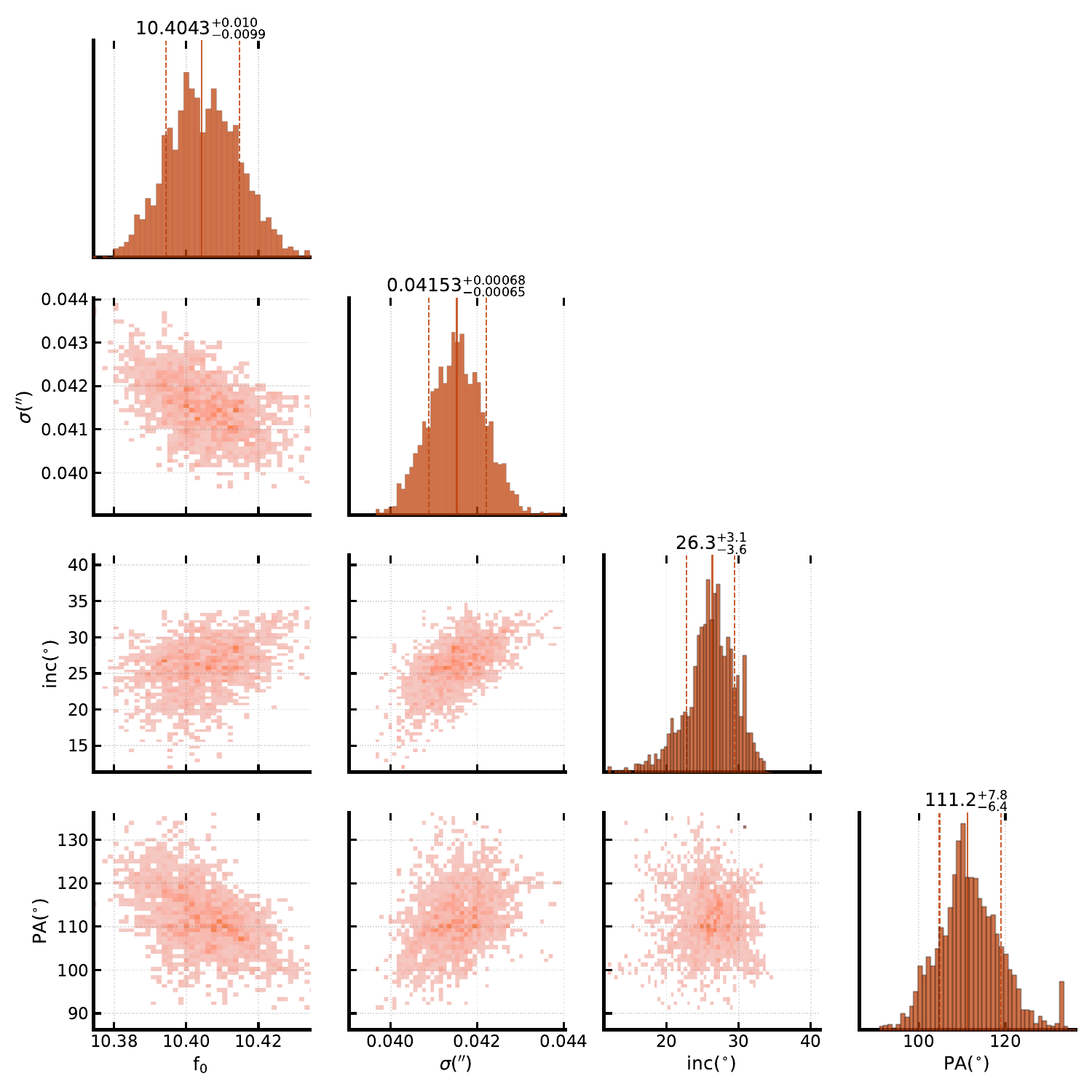} 
\caption{Corner plot of the MCMC performed by \texttt{galario} of the ALMA Band 6 continuum visibilities retrieving the disk parameters.}
\label{Fig: galario}
\end{figure} 
After performing self-calibration to the ALMA datasets, we used the continuum visibilities tables for each ALMA band to recover the disk's geometrical properties and axisymmetric radial brightness profiles.  Due to the unknown geometric properties of the disk, we fitted the ALMA visibilities with the \texttt{galario} package and a gaussian profile. We used the ALMA 1.3 mm continuum flux measurement and continuum image to constrain the peak and width of the gaussian, while we used a broad range for the inclination and position angle of the disk. Then, we show the resulting corner plot after the fitting in Figure \ref{Fig: galario}. The medial values for inclination and PA are reported in Table \ref{Tab: EKCHa_pars}. We used the \texttt{frank} package \citep{Jennings} that fits the data through a non-parametric approach but fixing the geometric properties and phase offset, using the ones obtained from \texttt{galario}. Given the low resolution of the data and small disk size (see Table \ref{Tab: Beam}), we set a coarse radial grid with 10 cells of 3 mas each fitting a total radius of 30 mas. For the \texttt{frank} fitting we used $(\alpha, w_{\rm smooth})=(1.2, 10^{-2})$.

\begin{figure}[h]
    \includegraphics[width=0.5\textwidth]{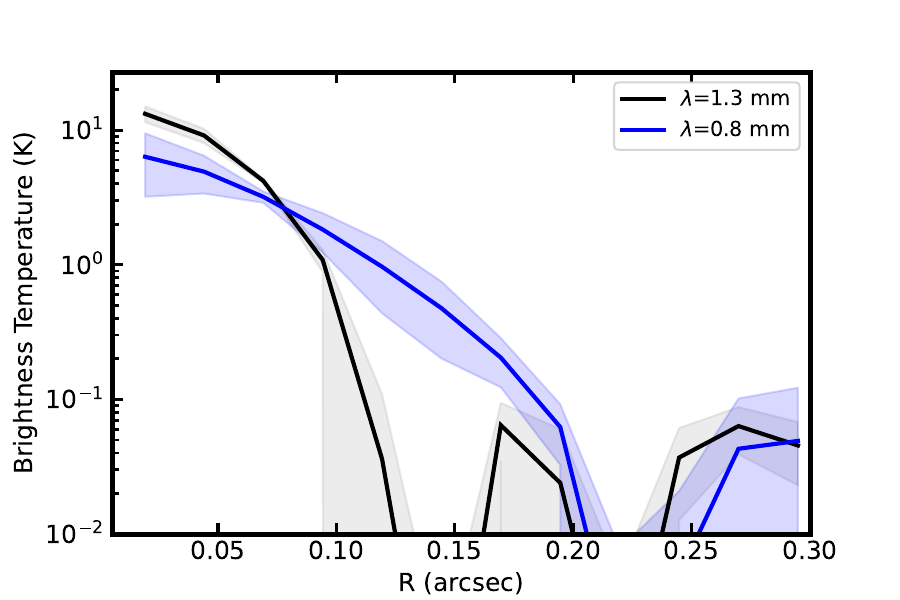} 
\caption{Frank radial profiles of the ALMA continuum emission after bootstrapping with 200 iterations.}
\label{Fig: frank}
\end{figure} 

After recovering for the disk geometry, we fitted for the radial profile with 2000 steps, or until reaching convergence, defined with a tolerance value lower than 10${-4}$. To measure uncertainties for our retrieved radial profiles, we performed a bootstrapping with 200 iterations. Then, we used the 16th and 84th quantiles of the bootstrapped profiles as uncertainties. The radial profiles are shown in Figure \ref{Fig: frank} with the shaded area indicating the confidence intervals from the bootstrapping analysis.


\end{appendix}

\end{document}